%% file: lcwsHiggsSummary_arXiv.tex
\documentclass[twoside]{LCWS11}
\usepackage[latin1]{inputenc}
\usepackage[dvips]{graphicx,epsfig,color}
\usepackage{wrapfig,rotating}
\usepackage{amssymb,amsmath,array}
\usepackage{cite}
\include{paperdef}
\graphicspath{{figs/}}

\begin{document}

\include{lcwsHiggsSummary_titlepage}

\include{lcwsHiggsSummary_main}

\end{document}

%% file: lcwsHiggsSummary_titlepage.tex
\thispagestyle{empty}
\setcounter{page}{0}
\def\thefootnote{\fnsymbol{footnote}}

\begin{flushright}
\mbox{}
\end{flushright}

\vspace{1cm}

\begin{center}

{\large\sc {\bf \input{lcwsHiggsSummary_title}}}
\footnote{Invited plenary talk given at the {\em LCWS 2011}, 
September 2011, Granada, Spain}

\vspace{1cm}

{\sc 
S.~Heinemeyer
\footnote{
email: Sven.Heinemeyer@cern.ch}%
}

\vspace*{1cm}

{\it
Instituto de F\'isica de Cantabria (CSIC-UC), 
Santander,  Spain 

}
\end{center}

\vspace*{0.2cm}

\BC {\bf Abstract} \EC
\input{lcwsHiggsSummary_abstract}

\def\thefootnote{\arabic{footnote}}
\setcounter{footnote}{0}

\newpage


%% file: lcwsHiggsSummary_title.tex
Progress for Higgs Bosons Physics at the LC

%% file: lcwsHiggsSummary_abstract.tex
A linear $e^+e^-$ collider (LC) could go into operation in
the next decade. The LHC is currently exploring the Higgs sector of the
SM, various supersymmetric extensions and other models. The LC is
necessary to complete the profile of a Higgs boson of any model. 
Experimental
analyses and theory calculations for Higgs physics at the LC are
currently performed. We review recent progress, as presented at the
{\em LCWS 2011} in Granada, Spain.

%% file: lcwsHiggsSummary_main.tex
\title{\input{lcwsHiggsSummary_title}}
\author{S.~Heinemeyer
\vspace{.3cm}\\
Instituto de F\'isica de Cantabria (CSIC)
E-39005 Santander, Spain
}

\maketitle

\begin{abstract}
\input{lcwsHiggsSummary_abstract}
\end{abstract}

\section{Introduction}

There is a world-wide consensus that a 
linear $e^+e^-$ Collider (LC) should be the next major project
in the field of high-energy physics. Depending on decisions on design
and site it could go into operation somewhen in the next decade. 
Currently the LHC is investigating the mechanism of electroweak
symmetry breaking, where the Higgs sector of the Standard Model (SM) and of
the Minimal Supersymmetric Standard Model (MSSM) are the leading candidates.
Consequently, an LC will 
take data several years after the LHC Higgs sector exploration. 
However, the physics case for the LC, especially in Higgs physics,
independent of what the LHC will find, has been made many
times~\cite{teslatdr,ilc,clic}. The complementarity
and the synergy of the two colliders and combined physics analyses has
been discussed extensively in \citere{lhcilc}.

A very important consideration in respect of the LC physics case is
the question what the LC can add to the LHC (or what a combined
LC/LHC analysis can add to the LHC). It has been
shown~\cite{teslatdr,ilc,clic,lhcilc,Snowmass05Higgs} that in {\em all}
conceivable scenarios of new physics
the LC can add valuable and important
information. In particular, it can add precision analyses,
pinning down model parameters extremely precisely. Moreover, in
contrast to the LHC, this can often be done in a model-independent
way. Furthermore in many scenarios the LC can discover new states that
cannot be detected at the LHC. The combination of these three
capabilities (precision measurements, model independent analyses,
discovery of new particles) enables the LC to determine the
underlying physics model. 

While the physics case for the LC has been made and the physics
potential of the LC has been analyzed, there are still many tasks
that have to be performed until the full potential of the LC can be
exhausted. This concerns the experimental analyses as well as the
(corresponding) theoretical calculations. In many scenarios the
feasibility of the experimental analyses has to be worked out in
detail. Theory calculations at the level of the anticipated LC
precision still have to be performed.
Progress in both directions has to be made over the next years in
order to be ready once the LC operation starts. A status of the field
and about recent progress was given at the {\em LCWS 2011} in Granada,
Spain. Here we briefly review the presentations about new experimental
analyses and new theory calculations given at the {\em LCWS 2011} in the
field of Higgs physics.


\section{Progress in Granada}
\label{sec:higgs}

If a (SM-like) Higgs mechanism is realized in nature, 
the LHC will find a Higgs boson and measure it
characteristics~\cite{atlas,cms,HcoupLHCSM,HcoupSFitter}. 
To be certain the state observed is indeed the Higgs boson, it is 
necessary to measure the couplings of this state to
the $W$~and $Z$~gauge bosons, and to fermions such as the top and
bottom quarks and the tau leptons. 
Consequently, the measurements at the LHC
include a mass determination at the per-cent level and coupling
constant determination at the level of 
10-20\%. However, in order to do this several assumptions about the
realization of 
the Higgs mechanism have to be made. Analyses could become much more
involved if the Higgs boson decay rates are strongly
different from the SM rates. Interesting
physics could easily hide in the 10-20\% precision achievable for the
Higgs boson couplings. Higgs self-couplings are extremely
complicated if not impossible to measure at the LHC. On the other
hand, all these problems can be overcome with the LC
measurements~\cite{teslatdr,clic,Snowmass05Higgs}.

The main points on the progress that will be necessary to fully exploit
the LC capabilities are:
\begin{itemize}
\item 
more analyses with full simulations of the relevant LC processes have to
be performed, \\[-1.5em] 
\item
higher precision in the theory calculations for the relevant processes
are needed to match the anticipated LC accuracy, \\[-1.5em]
\item
the Higgs/EWSB sector of more models has to be worked out,
\item
tools (encoding new models and/or high-precision calculations) have to become
available, \\[-1.5em] 
\item
the LHC/LC interplay has to be worked out in more detail. \\[-1.5em]
\end{itemize}
These issues have (partially) been addressed at the {\em LCWS 2011}. 
Progress has been reported e.g.\ about the following subjects (more details
can be found in the original publications).

\subsection*{Theory:}

\begin{itemize}

\item
Progress in loop calculations for the LC was summarized in
\citere{reuter}.

\item
Precision 2HDM studies for the LC~\cite{haber}.

\item
Theory calculations in the complex MSSM (cMSSM) for the Higgs production
from SUSY decays~\cite{heinemeyer} (to be implemented into the publicly
available code 
{\tt FeynHiggs}~\cite{feynhiggs}). 

\item
Higgs analyses for models with composite Higgs bosons~\cite{grojean}.

\item
Higgs potential and Higgs decays in triplet Higgs models~\cite{arhrib}.

\item
ILC analyses for models with Higgs triplets~\cite{yagyu}.

\item
Extended SUSY Higgs sectors and their decoupling properties at
colliders~\cite{kanemura}. 

\item
Theory evaluations for the Higgs sector of the Non Minimal Flavor
Violating MSSM~\cite{arana}. 

\item
Theory evaluations for the Higgs sector of the MSSM with heavy Majorana
neutrinos/sneutrinos~\cite{penaranda}. 

\item
Higgs production in models with universal extra
dimensions~\cite{oda}. 

\item
Predictions for a fermiophobic Higgs at the LC~\cite{gabrielli}.

\item
Multi tau lepton signatures in leptophilic two Higgs doublet
model at the LC~\cite{tsumura}.

\item
Determining the CP parity of Higgs bosons at the ILC
in the tau decay channels~\cite{berge}.

\item
Theory evaluations for the Higgs sector of a pure $B-L$
model~\cite{pruna}.

\item
Theory evaluations concerning dynamical symmetry breaking in SUSY
extensions of the Nambu Jona-Lasinio model~\cite{faisel}.

\end{itemize}

\subsection*{Experiment:}

\begin{itemize}
\item
Analysis for Higgs branching ratio measurements at the ILC~\cite{hiroaki}.

\item
Measurement of the top Yukawa coupling at the ILC~\cite{tanabe}.

\item
Improved $ZHH$ studies with ILD full simulation~\cite{suehara}.

\item
Analysis of light Higgs decays at CLIC~\cite{strube}.

\item
Analysis of heavy Higgs bosons at CLIC~\cite{battaglia}.

\end{itemize}


\section{Outlook}

With the results on Higgs boson searches published recently by
ATLAS~\cite{ATLASdec13} and CMS~\cite{CMSdec13} the prospects for Higgs
boson physics are brighter than ever before. 
It may well be that with the additional data to be collected in 2012 the
hints reported in \citeres{ATLASdec13,CMSdec13} will turn into a discovery of
a new state, compatible with a SM-like (or non-SM-like) Higgs particle,
with a mass around $\sim 125 \gev$. 
The particle physics community must be prepared for this possibility.

There is a large consensus that an LC running at relatively low energy (to
maximize the production cross section) is the best machine to study such
a new particle in detail. It would be the ideal motivation for the
construction of the LC. Only an LC could measure the profile of such a
Higgs-like state to a very high precision. This includes measurements of
its mass, its couplings to SM fermions and gauge bosons, its width as
well as its quantum numbers and spin. Possibly the high precision could
even help to distinguish various models predicting such a state from
each other. In the case of SUSY, an LC could strongly improve the reach
for heavier Higgs bosons, either in $e^+e^-$ pair production or in
$\ga\ga$ induced single production. 
However, we need more preparation from the theoretical and from the
experimental side to be ready to produce and exploit the necessary high
precision in LC Higgs physics. The contributions presented at the 
{\em LCWS 2011} presented a small, but relevant step into this direction.

In the case of a confirmation of the hints seen at $125 \gev$ a staged
approach could turn out to be an ideal solution. In the first stage at
low energy the LC can be used as a Higgs factory. In a second stage at
energies above $\sim 350 \gev$ it would turn additionally into a top
factory. Finally, in the highest energy stage the LC could continue the
TeV~scale exploration currently performed at the LHC. The various
options provided by an LC (GigaZ~\cite{gigaz}, the $\ga\ga$
mode, \ldots) have the potential to further improve our knowledge on the
Higgs sector (and any other sector) of the underlying model.





\begin{footnotesize}


\end{footnotesize}